\documentclass{ws-ijmpd}
\usepackage{amsmath,amssymb,hyperref,comment}

\begin{document}

\title{Gravitoelectromagnetism and stellar orbits in galaxies}

\author{Viktor T. Toth$^{\ddag}$\\~\\}
\address{$^\ddag$
Ottawa, Ontario K1N 9H5, Canada\\vttoth@vttoth.com}

\maketitle

\begin{history}
{Submitted~\today\par}
\end{history}

\begin{abstract}
Beyond the Newtonian approximation, gravitational fields in general relativity can be described using a formalism known as gravitoelectromagnetism. In this formalism a vector potential, the gravitomagnetic potential, arises as a result of moving masses, in strong analogy with the magnetic force due to moving charges in Maxwell's theory. Gravitomagnetism can affect orbits in the gravitational field of a massive, rotating body. This raises the possibility that gravitomagnetism may serve as the dominant physics behind the anomalous rotation curves of spiral galaxies, eliminating the need for dark matter. In this essay, we methodically work out the magnitude of the gravitomagnetic equivalent of the Lorentz force and apply the result to the Milky Way. We find that the resulting contribution is too small to produce an observable effect on these orbits. We also investigate the impact of cosmological boundary conditions on the result and find that these, too, are negligible.
\end{abstract}

\ccode{PACS numbers: 04.20.-q,04.30.-w,04.80.Nn}
~\par

The formalism known as gravitoelectromagnetism arises from a simple approximation: a solution of Einstein's field equations in the presence of slowly moving matter. This solution can be written in a very general form, valid to ${\cal O}(v^4/c^4)$, as\cite{Weinberg1972,Padmanabhan2010}:
\begin{align}
ds^2=\left(1+2\frac{\Phi}{c^2}\right)c^2dt^2-2\left(\frac{{\bf A}}{c}\cdot d{\bf r}\right)c~dt-\left(1-2\frac{\Phi}{c^2}\right)d{\bf r}^2,\label{eq:metric}
\end{align}
where $\Phi$ is the Newtonian gravitational potential, ${\bf A}$ is the gravitomagnetic vector potential and where we used $d{\bf r}^2=dr^2+r^2d\Omega^2$, e.g., in spherical coordinates. This form of the metric works for any perfect fluid matter distribution so long as it is nonrelativistic and slowly moving.

The advantage of this formalism is that it leads to test particle equations of motion that are already very familiar from classical electromagnetism. To see this, we first introduce the gravitomagnetic vector potential that corresponds to ${\bf A}$, in the form
\begin{align}
{\bf B}=\nabla\times {\bf A}.\label{eq:B}
\end{align}
To complete the analogy with electromagnetism, we can also define the gravitoelectric vector potential in the form
\begin{align}
{\bf E}=\nabla\Phi-\partial_t{\bf A}.
\end{align}
With these definitions at hand, we can write down the gravitoelectromagnetic equivalent of the Lorentz-force, acting on a particle with mass $m$, moving with velocity ${\bf v}$:
\begin{align}
{\bf F}=-m{\bf E}-m{\bf v}\times{\bf B}.\label{eq:F}
\end{align}
This formalism is of course well known and explained in detail in many introductory texts on general relativity. We note that, as expected, the formalism obeys the weak principle of equivalence: As ${\bf F}=m{\bf a}$ where ${\bf a}$ is the acceleration of the particle, the mass $m$ cancels out of Eq.~(\ref{eq:F}), leaving us with
\begin{align}
{\bf a}=-{\bf E}-{\bf v}\times {\bf B}.\label{eq:a}
\end{align}
Thus, all test particles follow similar trajectories regardless of their mass. Equivalently, we conclude that the gravitational charge-to-mass ratio (i.e., the ratio of gravitational mass to inertial mass) is the same for all material particles.

The contribution of the gravitomagnetic potential ${\bf B}$ to the equations of motion is responsible for the effect known as Lense--Thirring precession\cite{LenseThirring1918,Mashhoon1984}: the precession of a gyroscope orbiting a rotating mass, such as the Earth, experimentally verified by the {\em Gravity Probe B} experiment\footnote{\url{https://einstein.stanford.edu/}}, despite considerable challenges due to initially unmodeled noise.

In addition to gyroscopic precession, gravitomagnetism can also alter the orbits of test particles around a massive, rotating body, because of the extra contribution to ${\bf a}$ in Eq.~(\ref{eq:a}). This contribution is not only very small in the case of a spacecraft orbiting the Earth, it also cannot be readily decoupled from the quadrupole characterizing the gravitational field\cite{AsadaKasai2000}.

However, considering that a galaxy is immensely more massive than our planet, it is conceivable that this force plays a bigger role in stellar orbits in a spiral galaxy such as our Milky Way. This is potentially significant, since the rotation of spiral galaxies deviate noticeably from Newtonian predictions assuming only the presence of ``ordinary'' (i.e., baryonic) matter. Actual rotation curves are ``flat'' (instead of showing the expected $\propto r^{-1/2}$ dropoff that would naively follow for Keplerian orbits around a central mass) and the angular velocity significantly exceeds the Newtonian circular orbital velocity associated with the estimated baryonic mass of the galaxy. While the flatness of the rotation curve may be addressed, in part, by suitable, physically reasonable bulge-and-disk models, the high rate of rotation is not compatible with the observed (stellar and gas) galactic mass.

The now conventional explanation for these anomalous galaxy rotation curves is the assumed presence of a hypothetical halo of dark matter. The mass of this halo is ${\cal O}(10)$ higher than the observed baryonic mass, and the halo extends beyond the visible boundaries of the host galaxy. With these properties, a dark matter halo model can replicate the observed rate of rotation. However, despite numerous attempts to detect it, we have so far been unable to ascertain the existence of dark matter independently. This leaves open the possibility that perhaps dark matter doesn't even exist: that the observed rotation curves are due to the nature of gravity itself, either in the form of a modification of Einstein's theory, or in the form of a hitherto underappreciated aspect of the theory.

This is where gravitomagnetism enters the picture, through the intriguing possibility that the resulting gravitational Lorentz-force can account for the anomalous rotation curves of galaxies. To investigate this possibility, we first note that for a compact source of mass $M$ and angular momentum ${\bf J}$ at the origin, the gravitomagnetic potential has the specific form
\begin{align}
\Phi&=-\frac{GM}{r},\\
{\bf A}&=-\frac{2G}{c^2}\frac{{\bf J}\times{\bf r}}{r^3}.
\end{align}

In the case of a stationary gravitational field and circular motion, the magnitude of the radial acceleration due to the gravitomagnetic term can therefore be estimated easily:
\begin{align}
a_B=-v|{\bf B}|=-\frac{4G}{c^2}\frac{J~v}{r^3}.
\end{align}
So how does this work out here in the Milky Way? To get a crude estimate of the Milky Way's angular momentum, we can assume $M_\star=10^{11}M_\odot$ and a characteristic radius of $r=8$~kpc, which happens to be the approximate distance of the Sun from the central bulge. The orbital speed of the solar system is $\sim 200$~km/s. Multiplying these together, we obtain
\begin{align}
J_\star=M_\star vr\sim 10^{67}~{\rm J}\cdot{\rm s}.
\end{align}
This value is also consistent with published estimates.

Correspondingly, for the magnitude of the gravitomagnetic vector potential, we obtain
\begin{align}
B\sim 2\times 10^{-21}~s^{-1}.
\end{align}
This results in the radial acceleration
\begin{align}
a_B=v~B\sim 4\times 10^{-16}~{\rm m}/{\rm s}^2.
\end{align}
In contrast, the centrifugal acceleration of the solar system as it orbits the Milky Way is given by
\begin{align}
a_\odot=\frac{v^2}{r}\sim 1.6\times 10^{-10}~{\rm m}/{\rm s}^2.
\end{align}
We can see, then, that the gravitomagnetic effect is quite negligible, as it is more than five orders of magnitude smaller than the Newtonian value of the centripetal acceleration corresponding to the orbit of the solar system.

There is one additional possible point that might be raised. In Eq. (\ref{eq:metric}) we implicitly assumed that spacetime is asymptotically flat, described by the Minkowski metric at great distances. This is obviously not true in our expanding, matter-filled universe. But can the effects of this deviation from a Minkowski background be significant on the scale of a single galaxy?

To answer this question, we consider another metric: McVittie's metric\cite{McVittie1933,SKMHH2003} for a compact, spherically symmetric source of gravitation, with mass $M$ embedded in a Friedmann--Lema\^itre--Robertson--Walker (FLRW) cosmological background. In comoving coordinates, McVittie's metric is given by
\begin{align}
ds^2 &= \frac{\left[1 - \dfrac{GM}{2c^2aR}\right]^2}{\left[1 + \dfrac{GM}{2c^2aR}\right]^2}~c^2dt^2 - a^2\left[1 + \frac{GM}{2c^2aR}\right]^4 (dR^2 + R^2 d\Omega^2),
\end{align}
where $a=a(t)$ is the dimensionless scale factor of the FLRW universe. Switching to coordinates that employ a length scale that is not a function of time (i.e., no expanding meter sticks), in the form
\begin{align}
r = aR,
\end{align}
we obtain
\begin{align}
a~dR &= dr - Hr~dt,
\end{align}
where $H=\dot{a}/a$. Then,
\begin{align}
ds^2 &= \frac{\left(1 - \dfrac{GM}{2c^2r}\right)^2}{\left(1 + \dfrac{GM}{2c^2r}\right)^2}~c^2dt^2 - \left(1 + \frac{GM}{2c^2r}\right)^4 [(dr - Hr~dt)^2 + r^2d\Omega^2 ].
\end{align}
Of course, $\Phi=-GM/r$ and in the limit $\Phi\ll c^2$ and $Hr\ll c$, we recover $ds^2$ in the following approximate form:
\begin{align}
ds^2 &= \left(1 + 2\frac{\Phi}{c^2}-\frac{H^2r^2}{c^2}\right)c^2dt^2+2\frac{Hr}{c}dr~ c~dt- \left(1 - 2\frac{\Phi}{c^2}\right) d{\bf r}^2.
\end{align}

Comparing against (\ref{eq:metric}) we can thus directly estimate the cosmological contribution. For the Newtonian scalar gravitational potential, this contribution is given by $(Hr/c)^2$. Using $H\sim 70$~km/s/Mpc and $r=8$~kpc as characteristic values (the latter being the approximate distance of the Sun from the central region of the Milky Way), we have $(Hr/c)^2\sim 3.5\times 10^{-12}$, which is negligible in comparison to $\Phi/c^2=GM_\star/c^2r\sim 6\times 10^{-7}$.

The contribution to the gravitomagnetic term, however, is larger. Its dimensionless magnitude is given by $Hr/c\sim 1.9\times 10^{-6}$ for the Milky Way, which means that it exceeds $|{\bf A}|/c$ by more than three orders of magnitude. The term $r~dr$, however, a purely radial term: its curl, hence the corresponding contribution to ${\bf B}$ in Eq.~(\ref{eq:B}), therefore both vanish.

This demonstrates explicitly that any cosmological contribution to the orbits of stars is negligible and can be safely ignored when we consider the gravitomagnetic potential of a typical spiral galaxy, such as our Milky Way.

We can thus conclude, through explicit calculation, that neither the gravitomagnetic potential, nor cosmological boundary conditions have a measurable impact on galaxy rotation curves. Whatever the reason is for the anomalous rotation curves of spiral galaxies, it has to be sought elsewhere.

\section*{Acknowledgement}

VTT acknowledges the generous support of Plamen Vasilev and other Patreon patrons.

\bibliographystyle{ws-ijmpd}

\bibliography{refs}

\end{document}